\documentclass[prc,english,aps,twocolumn]{revtex4-1}
\usepackage[T1]{fontenc}
\usepackage[latin9]{inputenc}
\usepackage{amstext}
\usepackage{graphicx}
\usepackage{esint}
\usepackage{babel}
\begin{document}

\preprint{FTUAM-XX-2014}

\title{Dynamic versus static fission paths with realistic interactions}

\author{Samuel A. Giuliani}

\email{sam.and.giuliani@gmail.com}

\affiliation{Departamento de F\'\i sica Te\'orica, 
Universidad Aut\'onoma de Madrid, E-28049 Madrid, Spain}

\author{Luis M. Robledo}

\email{luis.robledo@uam.es}

\affiliation{Departamento de F\'\i sica Te\'orica, 
Universidad Aut\'onoma de Madrid, E-28049 Madrid, Spain}

\author{R. Rodr\'{\i}guez-Guzm\'an}

\email{raynerrobertorodriguez@gmail.com}

\affiliation{Department of  Physics and Astronomy, Rice University, 
Houston, Texas 77005, USA}

\affiliation{Department of Chemistry, Rice University, Houston, Texas 77005, USA}


\begin{abstract}
The properties of dynamic (least action) fission paths are analyzed and compared to the
ones of the more traditional static (least energy) paths. Both the BCPM and
Gogny D1M
energy density functionals are used in the calculation of the HFB constrained
configurations providing the potential energy and collective inertias. The
action is computed as in the WKB method. A full variational search of the least-action
path over the complete variational space of HFB wave functions is cumbersome and
probably unnecessary if the relevant degrees of freedom are identified.
In this paper, we consider the particle number fluctuation degree of freedom that
explores the amount of pairing correlations in the wave function. For a 
given shape, the minimum action can be up a factor of three smaller than
the action computed for the minimum energy state with the same shape. The
impact of this reduction on the lifetimes is enormous and dramatically
improves the agreement with experimental data in the few examples considered.
\end{abstract}

\maketitle


\section{Introduction}


The description of fission in the atomic nucleus is the subject of a renewed interest as a
consequence of potential applications both at the level of fundamental 
science and applications \cite{Krappe.12,Poenaru.96,Fissiona,Fissionb,Fissionc,Brack.72}. 
The mechanism of fission involves a delicate balance between quantum
mechanical effects inherent to the quantum many body problem and the properties 
of the nuclear interaction that are not well understood yet. Deepening
our understanding of fission is therefore important to understand
both effects in the dynamics of the atomic nucleus. 
Fission is also relevant to 
other areas outside traditional nuclear physics like astrophysics 
where the understanding of the nucleosynthesis of heavy elements in
explosive galactic environments through the r-process is of great relevance. A deeper 
understanding of fission is also of interest in safe energy 
production with new generation nuclear reactors or in radioactive waste 
degradation.
The theoretical interpretation of fission is based on
two properties of the parent  nucleus: the evolution of the energy
as the system traverses from its ground state to scission and the inertia
associated to the collective motion along that path. To improve in the former,
properties of the fission energy landscape as the inner and outer barrier
heights have been considered as a physical constraint in the fitting protocol 
of several energy density functionals (EDFs) \cite{Berger.84,skm,UNEDF1,McDonnell.13}. 
On the other hand, an effort
to improve the description of the collective inertias is underway \cite{Yuldash.99,Baran.11}. 
The gross features of fission can be understood rather well  from a microscopic 
perspective using the mean field Hartree-Fock-Bogoliubov (HFB) theory 
\cite{Ring.80} along with effective phenomenological interactions of different kinds.
Therefore, it is not surprising the large amount of 
studies devoted to this subject with Skyrme  interactions \cite%
{Bender.03,Erler.12,Staszczak.12,McDonnell.13}, Gogny ones \cite%
{Berger.84,Egido.00,Warda.02,Delaroche.06,Dubrai.08,Martin.09,
Perez-Martin.09,Younnes.09,Warda.11,Rodriguez.14,Rodriguez.14b}, based on the relativistic mean field \cite
{Abusara.10,Lu.12,Afanasjev.13} or other kind of energy density functionals
recently proposed \cite{BCPM,Giuliani.13}.

The traditional paradigm in fission is to describe  
the path to the scission configuration  by looking at the minimum energy in 
multidimentional energy landscapes. Several deformation parameters associated 
to multipoles of different orders are routinely used in those calculations. 
Surprisingly the role played by pairing in the dynamical aspects of the theory 
has attracted little attention and only recently the uncertainties associated 
to that degree of freedom in the values of fission observables have been 
explored \cite{Giuliani.13,Rodriguez.14,Rodriguez.14b}. 
The static (minimum energy) description is an approximation to the more quantal
approach where the action of the collective degrees of freedom is the quantity
driving the dynamics (the dynamical description). Assuming a static description
simplifies the computational problem and in addition, it was argued \cite{Baran.78}
that both approaches gave equivalent results when shape degrees of freedom alone
were considered as collective degrees of freedom (see \cite{Sadhukhan.13} for a 
recent result).
On the other hand, that pairing is a fundamental ingredient for fission dynamics was already pointed out 
forty years ago by Moretto and Babinet \cite{Moretto.74}
in a seminal paper where the impact of pairing correlations on the
action was analyzed in a simple, yet realistic, model. The main conclusion
of that paper was that the competition between the collective inertia
decreasing as the square  of the inverse pairing gap \cite{Brack.72,Bertsch.91} and the energy
increasing as the square of the pairing gap lead to a minimum of the  action
at a pairing gap twice as large as the one of the minimum energy.
The importance of pairing vibrations in nuclear dynamics was empashized
by B\'es et al \cite{Bes.70} in the context of the cranking model and by 
Gozdz et al \cite{Gozdz.85,Staszczak.85} in the Generator Coordinate Method (GCM) framework.
The idea of coupling pairing vibrations to nuclear fission dynamics 
was pursued further by the Lublin school \cite{Staszczak.85} in a series of papers
where realistic fission calculations seeking for the minimum of the
action were performed -see \cite{Pom07} for a recent overview
and additional references. When pairing was included as dynamical variable
they observed reductions of a few orders of magnitude (up to 8 in Cm isotopes)
in the spontaneus fission lifetimes $t_\textrm{SF}$ as compared to the minimum energy results. 
In nuclei with higher and broader barriers (like the ones considered here) the
reduction is expected to be larger.

In this paper we study the predictions for spontaneous fission 
lifetimes when considering the minimum action scheme instead of the
minimum energy one. We will consider pairing as a relevant degree
of freedom in the parameter space used in the minimum action search.
Calculations with effective, yet realistic, interactions
including finite range ones (Gogny D1M \cite{Decharge.80,D1M}) and pure energy density functionals
like the BCPM functional \cite{BCPM} will be considered. We will restrict ourselves
to several isotopes of uranium where experimental data are available. The main 
reason for this choice is the large discrepancies observed between theory and experiment
and due to very high and wide fission barriers.


\section{Methods}


The path from the ground state to the scission point is assumed to be driven by the  least
action principle, where the action is computed as
\begin{equation}\label{Eq:S}
S = 2 \int_a^b ds \sqrt{2B(s)(V(s)-E_0)}.
\end{equation}
The variable $s$ is used to parametrize the fission path, and all the relevant 
collective degrees of freedom depend upon it. The inertia
is given by the general expression
\begin{equation}\label{Eq:B}
B(s) = \sum_{ij} B_{ij} \frac{dq_i}{ds} \frac{dq_j}{ds},
\end{equation}
where the $q_i(s)$ are the values along the fission path of the relevant degrees of freedom like the 
axial quadrupole ($Q_{20}$), octupole ($Q_{30}$) and hexadecapole ($Q_{40}$) moments and in our case the
particle number fluctuation $\Delta N^2$. The $B_{ij}$ are the collective inertias for
each pair of degrees of freedom and computed using the perturbative
cranking approximation \cite{Libert.99,Baran.11}. Both the expressions obtained in the 
Adiabatic Time Dependent HFB (ATDHFB)
or the  Generator Coordinate Method (GCM) approximations will be used.
The potential $V(s)$ is given by the HFB energy with corrections from
beyond mean field effects included. The most important is the
rotational energy correction $\epsilon_\textrm{ROT} (Q_{20})$ 
which is the correlation energy gained by
restoring angular momentum quantum numbers. It is proportional to
deformation and can reach several MeV. It therefore has the potential
to reduce the fission barrier heights substantially. It has been 
computed using a well performing  approximation to the exact quantity \cite{egi04}.
Less important for
fission is the zero point energy correction associated to quantum fluctuations
of the collective $Q_{20}$ degree of freedom $\epsilon_0 (Q_{20})$. Details
on how these quantities are evaluated are given in \cite{Giuliani.13}
They are computed with the constrained Hartree-Fock-Bogoliubov (HFB)
wave functions obtained with state-of-the-art semi-phenomenological energy
density functionals of the Gogny and BCP type. Axial symmetry is preserved
in the calculation but reflection symmetry is allowed to break at any stage of the
calculations. Finally, the computer code HFBaxial \cite{HFBaxial} has been used.

For this exploratory calculation we have made several simplifying assumptions:
\begin{enumerate}
	\item Separated proton and neutron pairing correlations are not considered.
	Instead, a single constraint in the total particle number fluctuation is
	considered. This simplification reduces the computational cost by a 
	factor of a few tens (depending on the number of $\langle \Delta N^2\rangle${}
	values considered).
	
	\item The fission path is parametrized by the mass quadrupole moment.
	As customary in fission calculations based in the minimum
	energy principle the coupling with other degrees of freedom in the
	evaluation of the inertia Eq. (\ref{Eq:B}) is neglected and only the
	quadrupole inertia is considered.
	
	\item As a consequence, the minimum action path does not require for
	its determination of sophisticated linear programming techniques. It reduces to a
	simple minimization for fixed quadrupole moments of the integrand in the action 
        considered as a function of the other degrees of freedom.

\end{enumerate}

The constrained HFB wave functions $|\varphi (Q_{20}, \langle \Delta N^2\rangle) \rangle $
are obtained using an approximate second order gradient method \cite{rob11} with the
HFBaxial code \cite{HFBaxial}. Two EDF will be used in the calculation to make sure
that the conclusions are independent of the details of the nuclear interaction. 
First we use the  Gogny \cite{Decharge.80} force with the 
D1M \cite{D1M} parametrization, whose main characteristic is its finite range that allows
a consistent treatment of pairing correlations using the same 
two body force that determines de particle-hole Hartree- Fock potential. Altough D1M was
specifically tailored to describe binding energies, a series of papers \cite{Rob.08,RRG.10,RRG.10b,Rob11b,Rodriguez.14,Rodriguez.14b} 
have shown that D1M provide good results for many low energy nuclear observables. 
The other EDF used is the recently proposed BCPM EDF \cite{BCPM} with its density dependent 
pairing and effective mass equal to one. Again BCPM has been fitted to reproduce binding energies
but has proved to produce consistent low energy results in a variety of cases \cite{BCPM}.


\section{Results}

\begin{figure*}[htb]
\includegraphics[width=0.3\textwidth]{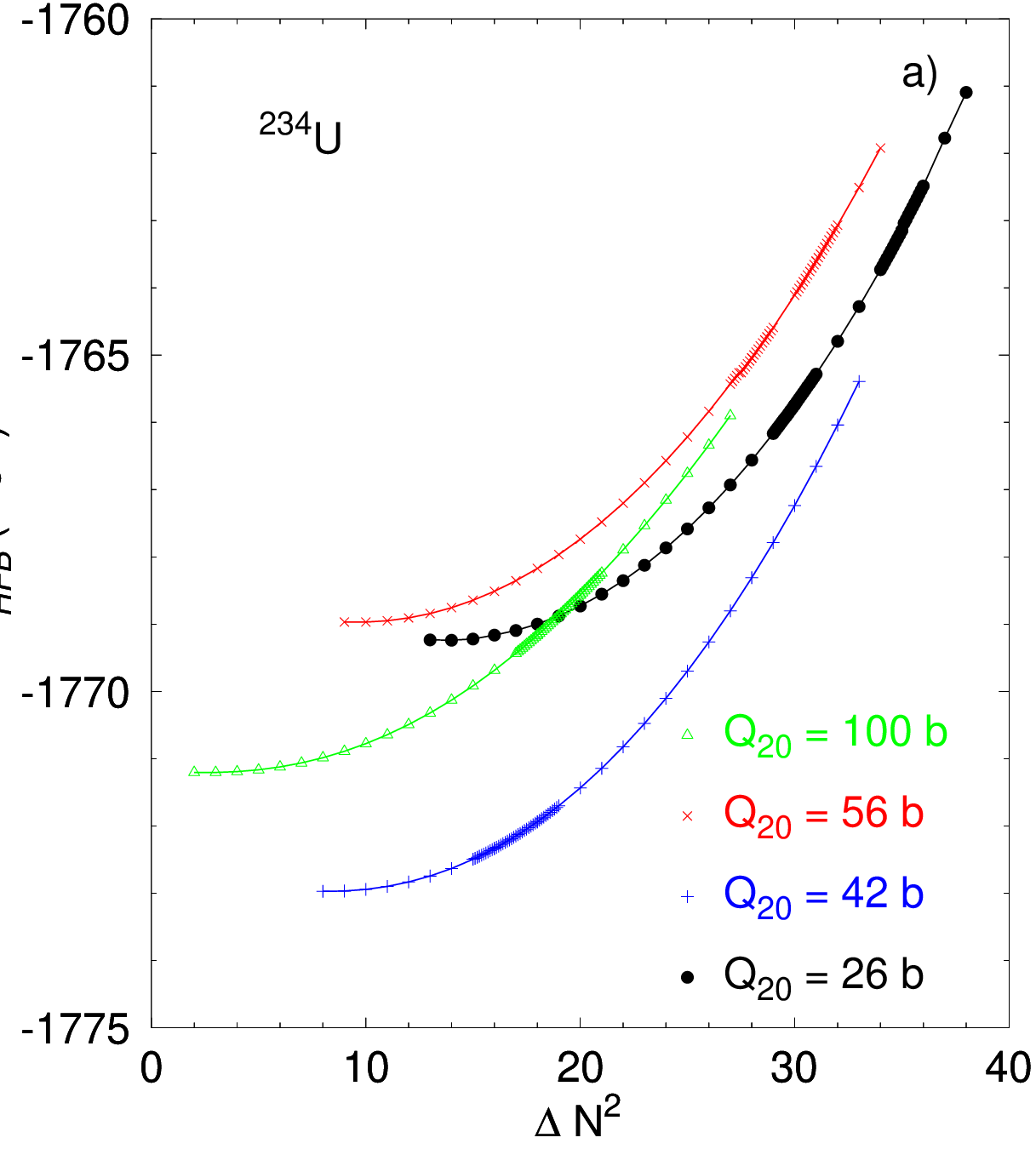}
\includegraphics[width=0.3\textwidth]{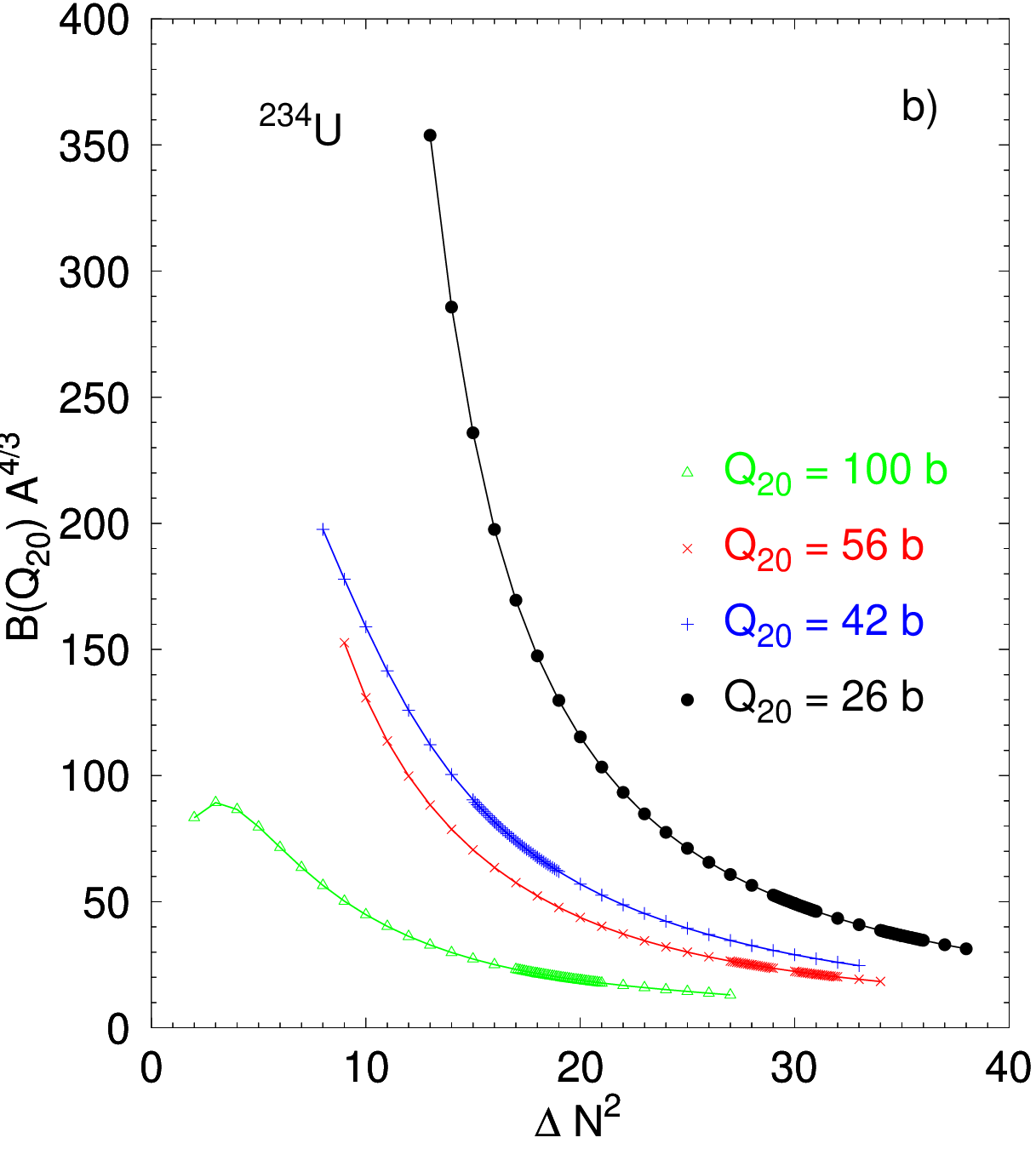}
\includegraphics[width=0.31\textwidth]{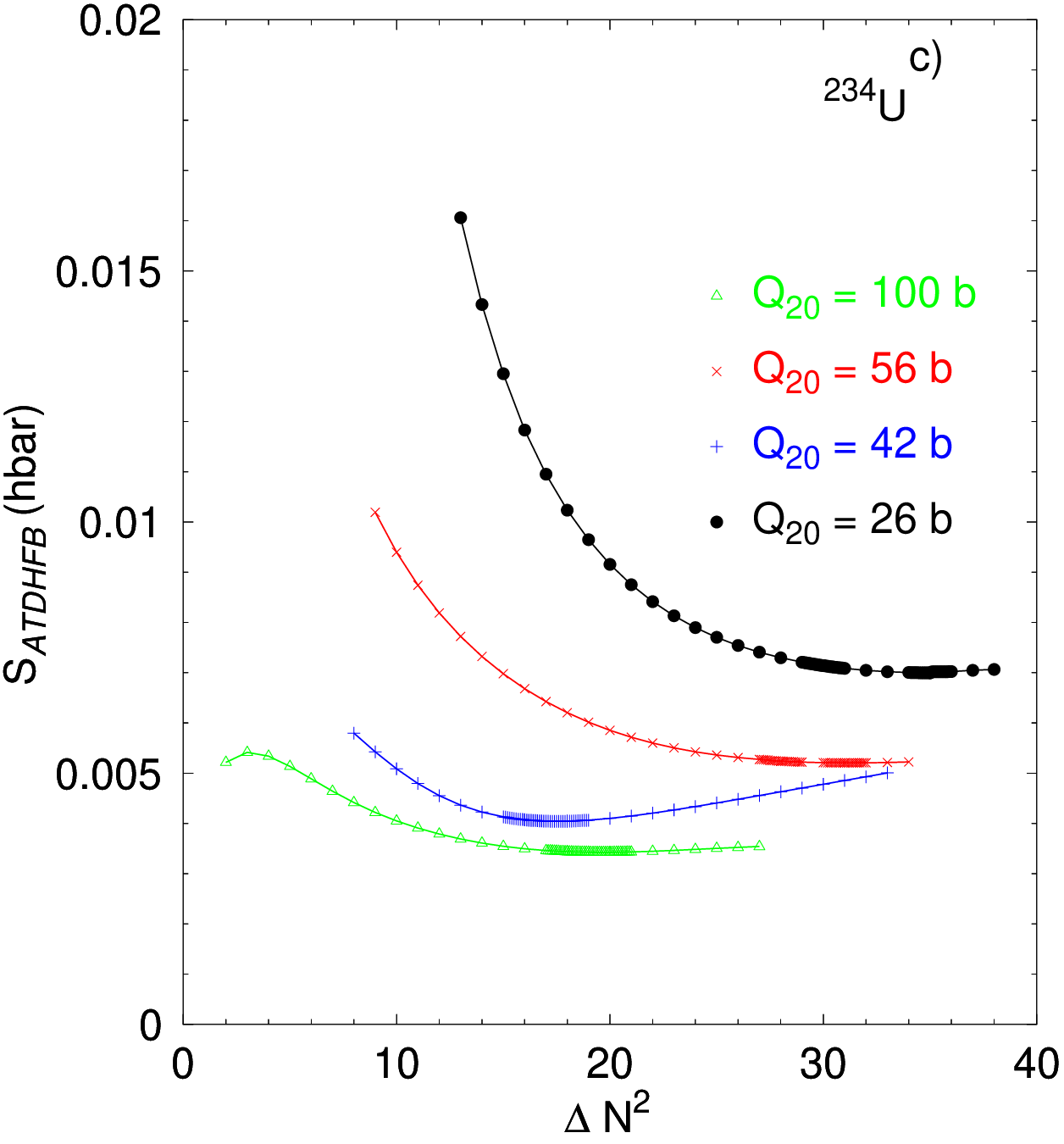}

\caption{The HFB energy, panel a); the collective inertia, panel b) and
the action, panel c) are represented as a function of
particle number fluctuation $\Delta N^2$ for several relevant 
$Q_{20}$ values (given in barns). \label{fig:Action}}
\end{figure*}
To illustrate the procedure we will focus on the nucleus $^{234}$U whose
fission properties were thoroughly studied in \cite{Giuliani.13} using 
the BCPM and Gogny D1M functionals. For both functionals, the nucleus 
presents its ground  state minimum at 
$Q_{20}=12$b. Also for both functionals, the first fission barrier is 
located at $Q_{20}=26$b, the fission isomer minimum is at $Q_{20}=42$b 
and the second fission barrier is at $Q_{20}=62$b. As discussed thoroughly 
in \cite{Giuliani.13} the main difference between the two
functionals is the inertia that is roughly three times larger for BCPM.
This is probably a consequence of its larger effective mass and the different
pairing interactions used.

To find the $\langle \Delta N^2\rangle$ value minimizing the action
we carry out constrained HFB calculations 
starting at the $\langle \Delta N^2\rangle$ value minimizing the energy for each $Q_{20}$ value of the fission path. 
In this way we obtain curves for the relevant physical quantities as a function of $\langle \Delta N^2\rangle$
for each value of $Q_{20}$. An example of such curves is shown in Fig. \ref{fig:Action} 
for the $Q_{20}$ values corresponding to the first and second fission barriers,
the fission isomer and at $Q_{20}=100$b as a characteristic value for very 
large elongations. We observe in panel a) an almost parabolic behavior for the energy as 
a function of $\langle \Delta N^2\rangle$ with the minimum located at the self-consistent solution.
In panel b) the decrease of the inertia with the inverse of $\langle \Delta N^2\rangle^2$
(corresponding to the law $B\approx 1/\Delta^2$ \cite{Brack.72,Bertsch.91}) is noticed. 
Finally, the integrand in the action $S$ of Eq. (\ref{Eq:B}),
computed with the ATDHFB version of the inertia is plotted in panel c). 
It shows a minimum at rather
large $\langle \Delta N^2\rangle$ values as compared with the ones of the 
minimum energy solution. The minimum value of the action integrand does 
not coincide with the value obtained (selconsistently) from the 
minimization of the energy, being the latter one up to a factor 
three larger than the true minimum. This large
quenching reduces considerably the action Eq. (\ref{Eq:B}) that appears
in the exponential of the WKB formula. Therefore, the impact on the
$t_\textrm{SF}$ values is large. In this case we obtain $0.18 \times 10^{23}$ s and 
$0.21 \times 10^{19}$ s depending on the choice of the collective mass, ATDHFB or GCM
respectively. Those values have to be compared with the ones obtained
minimizing the energy, namely $0.81 \times 10^{43}$ and $0.70 \times 10^{30}$ s. We observe
a reduction of 20 and 11 orders of magnitude that brings the theoretical
predictions in much closer agreement with experiment. Another beneficial
side effect of the action minimization is that results are much less sensitive to
the approach used to compute the inertias. This remarkable reduction was
observed in previous calculations \cite{Pom07} by the Lublin group in other
heavier nuclei. 

\begin{figure}
\includegraphics[width=0.8\columnwidth]{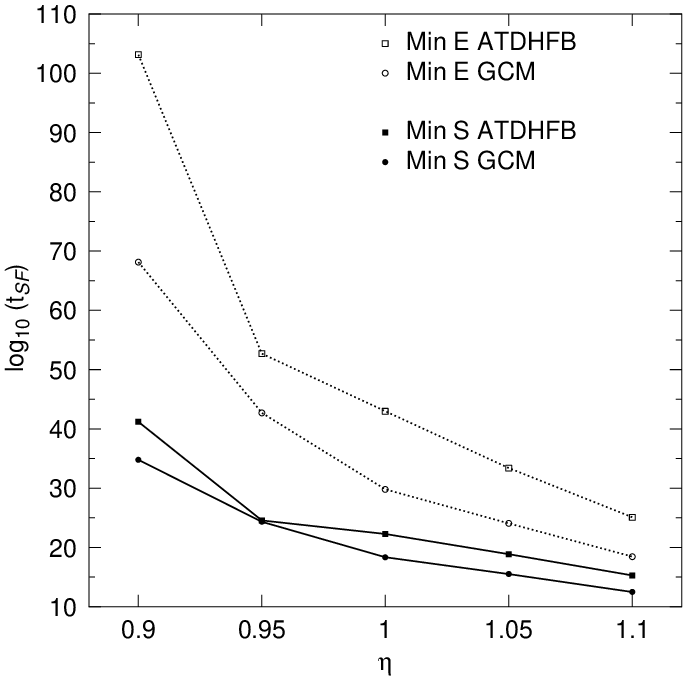}

\caption{Spontaneous fission lifetimes of $^{234}$U obtained with different approaches
as a function of the $\eta$ factor multiplying the pairing interaction.\label{fig:eta}}
\end{figure}

The minimum action solutions have $\langle \Delta N^{2} \rangle$ values much
larger than the minimum energy configurations and therefore have stronger
pairing correlations. In \cite{Giuliani.13,Rodriguez.14,Rodriguez.14b} we analyzed the
impact on $t_\textrm{SF}$ of increasing the pairing strength. We observed
a strong dependence of the results for $t_\textrm{SF}$ with small changes on the pairing
strengths. We have repeated those calculations but now computing the lifetimes
using the minimum action principle. The results obtained with BCPM \cite{Giuliani.13} (the
ones for Gogny D1M are similar \cite{Rodriguez.14}) are show in Fig. \ref{fig:eta} as a function
of the factor $\eta$ used to modify the pairing strength.

We observe that the strong dependence of $t_\textrm{SF}$ with $\eta$ in
the minimum energy calculation gets severely reduced in the minimum action
case. This is not so surprising as in the minimum action case we are already
exploring the amount of pairing correlations in the search for the minimum
action configuration. This search can compensate somehow the changes induced by
the varying interaction strength. It is also noteworthy to mention that now the ATDHFB and
GCM results are much closer than in the minimum energy case.

Recently, fission calculations based on the minimum action principle and using
one of the Skyrme variants have been presented. The variables considered are the axial $Q_{20}$ and 
triaxial $Q_{22}$ multipole moments. The main conclusion of that work
is that minimum energy and minimum action results are essentially the
same. Obviously, the result is not in contradiction with ours as different
degrees of freedom are considered. In order to further investigate this
aspect we have carried out action minimization calculations considering the
pairs ($Q_{20}$, $Q_{30}$) and  ($Q_{20}$, $Q_{40}$) of collective variables
to explore the minimum action trajectory. The results obtained with BCPM are summarized in Table \ref{table:I}.
When compared with the results obtained using pairing correlations as a 
relevant collective variable in the action, the ones obtained considering
$Q_{30}$ and $Q_{40}$ show a negligible impact being essentially the same
as the ones obtained minimizing the energy. This 
can be understood as a consequence of the weak dependence of the collective
inertia $B(Q_{20})$ with $Q_{30}$ and $Q_{40}$. The conclusion is
also evident, only the pairing degree of freedom is relevant for the action
minimization calculation due to the dependence of the inertia with the
inverse of the pairing gap.

\begin{table}
	\begin{tabular}{lcc}\hline\hline
	   Method                                & $t_\textrm{SF}$ (ATDHFB) & $t_\textrm{SF}$ (GCM) \\ \hline\hline
	   $E_\textrm{Min}$                      & $0.81 \times 10^{43}$           & $0.70 \times 10^{30}$        \\
	   $S_\textrm{Min}(Q_{20},\Delta N^{2})$ & $0.18 \times 10^{23}$           & $0.21 \times 10^{19}$        \\
	   $S_\textrm{Min}(Q_{20},Q_{30}) $      & $0.44 \times 10^{42}$           & $0.64 \times 10^{29}$        \\
	   $S_\textrm{Min}(Q_{20},Q_{40}) $      & $0.12 \times 10^{43}$           & $0.10 \times 10^{29}$        \\ \hline\hline
	 \end{tabular}
	 \caption{Spontaneous fission lifetimes computed with the ATDHFB and GCM approximations to
                  the collective inertias. Results for different sets of collective variables
                  used to search for the minimum of the action are given.\label{table:I}}
\end{table} 

Finally, we show in Table \ref{table:II} a summary of the $t_\textrm{SF}$
obtained minimizing the action and using the two EDF considered  and the two
variants of the collective inertia.  When compared with the results obtained
minimizing the energy \cite{Giuliani.13,Rodriguez.14,Rodriguez.14b}, the action minimization
results show a much better agreement with experiment as well as a much reduced
dispersion with the interaction used and the variant of the collective mass
considered. This is a very important result as the theoretical interpretation
of fission was hampered by the large variability on the $t_\textrm{SF}$
values depending on the pairing strength.

\begin{table}[tbh]
	\begin{tabular}{lccccc} \hline\hline
	          & \multicolumn{2}{c}{ Gogny D1M }& \multicolumn{2}{c}{ BCPM }& Exp \\	
            &  $S_{min}$ (A) & $S_{min}$ (G) &  $S_{min}$ (A) & $S_{min}$ (G) & \\ \hline\hline
$^{232}$U	& $ 5.4 \times 10^{20}$	& $ 1.5 \times 10^{17}$	&$ 8.6 \times 10^{19}$	&$ 8.2 \times 10^{16}$	& $2.5 \times 10^{21}$\\
$^{234}$U	& $ 7.3 \times 10^{21}$	& $ 2.7 \times 10^{18}$	&$ 1.7 \times 10^{22}$	&$ 2.1 \times 10^{18}$  & $4.7 \times 10^{23}$\\
$^{236}$U	& $ 2.8 \times 10^{23}$	& $ 9.9 \times 10^{19}$	&$ 1.9 \times 10^{22}$	&$ 1.2 \times 10^{18}$	& $7.8 \times 10^{23}$\\
$^{238}$U	& $ 1.6 \times 10^{24}$	& $ 6.7 \times 10^{20}$	&$ 6.1 \times 10^{21}$	&$ 6.2 \times 10^{17}$	& $2.6 \times 10^{23}$\\   \hline\hline
	 \end{tabular}
	 \caption{Spontaneous fission lifetimes for several uranium isotopes obtained using the minimum action principle 
                  with the ATDHFB (A) and GCM (G) variants of the collective inertias.\label{table:II}}
\end{table}

\section{Conclusions}

The main conclusion of the paper is that the scheme used to compute 
$t_\textrm{SF}$  where the minimum action principle is used and  
the amount of pairing correlations is considered as collective variable 
gives results that strongly differ from the one where the energy is minimized
to determine the fission path. The spontaneous fission lifetimes computed 
minimizing the action are several orders of magnitude smaller than the ones 
obtained from the traditional minimization of the energy. This decrease 
improves dramatically the agreement with the experimental data. In addition,
the minimum action results show a weaker dependence with the ingredients of the
calculation than the minimum energy ones. This is an important result for the
credibility of mean field techniques to reproduce fission dynamics observables.

It is also shown that considering other variables in the minimization 
of the action like the multipole moments $Q_{30}$ and $Q_{40}$ has little
impact on the results being equivalent to the energy minimization ones.

\begin{acknowledgments}
Work supported in part by MICINN grants Nos. FPA2012-34694, FIS2012-34479
and by the Consolider-Ingenio 2010 program MULTIDARK CSD2009-00064. 
This work was initiated while one of the authors 
(LMR) participated at the INT13-3 program. The warm hospitality of 
the Institute for Nuclear Theory and the University of Washington is 
greatly acknowledged.
\end{acknowledgments}

\end{document}